\documentclass[prl,aps,twocolumn,superscriptaddress,longbibliography,floatfix]{revtex4-1}

\usepackage[
  colorlinks=true,
  urlcolor=blue,
  linkcolor=blue,
  citecolor=blue,
  bookmarks=false,
  pdftitle={},
  pdfauthor={E. Cornfeld}
]{hyperref}

\usepackage{import}
\usepackage{subfiles}
\usepackage{graphicx}
\usepackage{bm}
\usepackage{amsmath}
\usepackage{amssymb}
\usepackage{mathtools}
\usepackage{amsfonts}
\usepackage[all]{xy}
\usepackage{xspace}
\usepackage{accents}
\usepackage{stmaryrd}
\usepackage{tabularx}
\usepackage{extarrows}
\usepackage{float}
\usepackage{multirow}
\usepackage{makecell}
\usepackage[dvipsnames]{xcolor}
\usepackage[normalem]{ulem}
\usepackage{pdfpages}

\makeatletter
\AtBeginDocument{\let\LS@rot\@undefined}
\makeatother

\newcommand{\ket}[1]{{\left|#1\right\rangle}}
\newcommand{\tket}[1]{{|#1\rangle}}
\newcommand{\bra}[1]{{\left\langle #1\right|}}

\newcommand{\Chiral}{\Pi}

\newcommand{\BZ}{\mathcal{BZ}}

\newcommand{\vk}{\mathbf{k}}
\newcommand{\vK}{\mathbf{k_\textbf{o}}}
\newcommand{\vKstar}{\mathbf{k}^{*}_\textbf{o}}
\newcommand{\vp}{\mathbf{p}}

\newcommand{\va}{\mathbf{a}}

\newcommand{\flag}[1]{\mathrm{Fl}_{#1}}

\newcommand{\Chern}{\mathrm{Ch}}
\newcommand{\Surf}{\mathcal{M}}

\newcommand{\Conn}{\mathcal{A}}
\newcommand{\Phase}{\gamma}
\newcommand{\Curve}{\Omega}

\newcommand\redsout{\bgroup\markoverwith{\textcolor{red}{\rule[0.5ex]{2pt}{0.4pt}}}\ULon}
\newcommand\bluesout{\bgroup\markoverwith{\textcolor{blue}{\rule[0.5ex]{2pt}{0.4pt}}}\ULon}

\newcommand{\ADhide}[1]{{}}

\newcommand{\SPHIDE}[1]{{}}

\newcommand{\mytitle}{Punctured-Chern topological invariants for semi-metallic bandstructures}

\begin{document}

\title{\mytitle}

\author{Ankur Das}
\email{ankur.das@weizmann.ac.il}
\affiliation{Department of Condensed Matter Physics, Weizmann Institute of Science, Rehovot 7610001, Israel}
\author{Eyal Cornfeld}
\affiliation{Department of Condensed Matter Physics, Weizmann Institute of Science, Rehovot 7610001, Israel}
\author{Sumiran Pujari}
\email{sumiran@phy.iitb.ac.in}
\affiliation{Department of Physics, Indian Institute of Technology Bombay, Mumbai, MH 400076, India}

\begin{abstract}
Topological insulator-based methods underpin the topological 
classification of gapped bands, including those surrounding semi-metallic 
nodal defects. However, multiple bands with gap-closing points can also
possess non-trivial topology.
We construct a general wavefunction-based ``punctured-Chern" invariant 
to capture such 
topology.
To show its general applicability, we analyze two 
systems with disparate gapless topology:
1) a recent two-dimensional fragile topological model 
to capture the various 
band-topological transitions and 2) a three-dimensional 
model with a triple-point nodal defect to
characterize its semi-metallic topology with \emph{half-integers}
that govern physical observables such as anomalous transport. 
This invariant also gives the classification
for Nexus triple-points ($\mathbb{Z}\times\mathbb{Z}$)
with certain symmetry restrictions, 
which is re-confirmed by abstract algebra.
\end{abstract}

\maketitle

Electronic band topology has emerged as an important 
area of investigation 
following the realization that topological insulators
are distinct than conventional band
insulators~\cite{Hasan_Kane_2010,Hasan_Moore_2011,Qi_Zhang_2011,Ando_2013,bernevig2013topological,Ludwig_2015,Stern_2016,Tokura_Yasuda_Tsukazaki_2019}.
Non-trivial topology of band structures arising from
the underlying wavefunction geometry 
drives this phenomenology. This has made the topological 
classification of bands
into an active field. This 
is done by formulating topological
invariants that distinguish different band topologies,
e.g. the familiar wavefunction-based Chern invariant~\cite{tknn}. 
They are generally described by different discrete groups
\cite{Altland_Zirnbauer_1997,Schnyder_etal_2008,Ryu_etal_2010,Kitaev_2009,Lu_Joannopoulos_Soljacic_2014,Cornfeld_Chapman_2019,Cornfeld_Carmeli_2021}.

Heretofore, such topological classification has 
primarily focused on 
insulating bands \cite{Schnyder_etal_2008,Ryu_etal_2010,Kitaev_2009,Hasan_Kane_2010}, 
e.g. the $\mathbb{Z}_2$ topological insulator \cite{Kane_Mele_2005_1,Kane_Mele_2005_2}.
Insulator-based classification is also used to characterize 
band degeneracies or nodes
in one higher dimension by classifying gapped bands that surround such a
node, e.g. Weyl points in $3d$ using $2d$ Chern classification
\cite{Vafek_Vishwanath_review_2014},
$2d$ Dirac points using $1d$ chiral insulator classification, etc. 
However, multiple bands may support non-trivial topology even in presence
of gap-closing. A recent
example is fragile topology that has arisen in the context of 
magic-angle twisted bilayer graphene~\cite{Po_etal_2018prx,Zou_etal_2018prb,Song_etal_2019prl,Po_etal_2019prb}.
It is understood as an obstruction to a localized 
Wannier representation that gets removed by addition of disconnected
topologically trivial bands~\cite{Po_Watanabe_Vishwanath_2018prl,Ahn_Park_Yang_2019prx,Bouhon_Black-Schaffer_Slager_2019prb,Bradlyn_etal_2019prb}.
Such
band topology  arises also in
presence of gap-closing points~\cite{Zou_etal_2018prb,Turner_Berg_Stern_2022prl}.
The  question then arises: how to characterize 
``gapless" topology, 
in particular by wavefunction-based invariants?

Recently, there has also emerged a new class of $3d$ semi-metals that host 
three-fold band degeneracies called Nexus triple-points~\cite{Heikkila_Volovik2015,Zhu_etal_2016}
where insulator-based methods again fail.
These three-fold nodal defects have been predicted in several 
candidates~\cite{Zhu_etal_2016,Weng_etal_2016,Weng_etal_2016_2,Hyart_Heikkila_2016,Chang_etal_2017,Zhang_etal_2017,Feng_etal_2018,Barman_etal_2019,Lenggenhager_etal_2021}
and seen in experiments on MoP~\cite{Lv_etal_2017},
WC~\cite{Ma_etal_2018}, and GeTe~\cite{Krempasky_etal_2021}.
The failure results from the doubly-degenerate lines
or nodal-lines  
that generically emanate from the triply-degenerate point 
(Fig.~\ref{fig:nexus_sketch})
and intersect with
any
enclosing surface necessarily, precluding surrounding gapped surfaces. 
The original papers 
(e.g. Refs.~\cite{Heikkila_Volovik2015,Zhu_etal_2016})
had focused on insulator-based $\mathbb{Z}_2$ invariants
on gapped $1d$ loops surrounding the nodal-lines
that gave the first inkling of the topology of Nexus defects. 
Another approach used
quaternionic charges associated with nodal-chains for $PT$-symmetric models along 
with mirror symmetry~\cite{Lenggenhager_etal_2021} based on earlier 
works~\cite{Ahn_etal_2018,Wu_Soluyanov_Bzdusek_2019, Bouhon_etal_2020} that has been 
recently consolidated further~\cite{Lenggenhager_etal_2021b,Lenggenhager_etal_2022}.
Both these approaches
relied on the insulator-based 
paradigm on gapped $1d$ loops. Yet another attempt~\cite{Das_Pujari_2020}
that focused instead on the surrounding $2d$ gapless surface consisted of a homological
characterization on a suitably extended manifold 
to account for analyticity everywhere.
Formulating a wavefunction-based invariant directly on the (extended)
manifold in one lower dimension surrounding the Nexus triple-point
-- similar in spirit to
the Chern invariant -- and its discrete group classification
is desirable but yet to be 
formulated within the general scope of semi-metallic band topology.

\begin{figure}[t]
    \centering
    \includegraphics[width=0.95\linewidth]{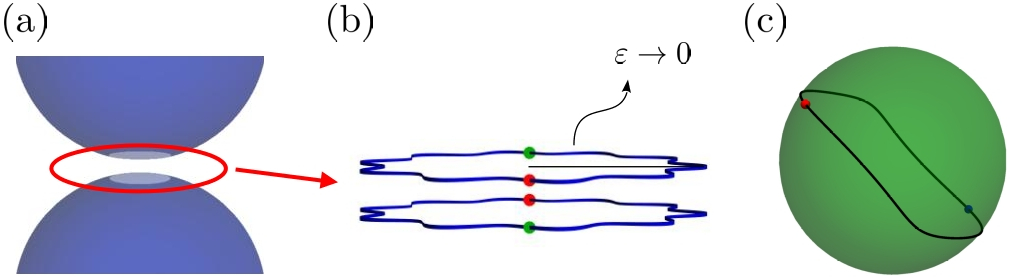}
    \caption{
    (a) Representation of the extended manifold near a puncture 
    due to the gap-closing between two bands~\cite{Das_Pujari_2020}. 
    (b) A zoom near the puncture on a piece of the manifold
    whose radius tends to zero.
    (c) Representation of the image
    on the $2d$ subspace (Bloch sphere) of this infinitesimal manifold. 
    It illustrates the antipodal nature of the image independent
    of the shape of the infinitesimal piece in panel (b)).}
    \label{fig:puncture_sketch}
\end{figure}

Motivated by the considerations above,
we construct below a wavefunction-based topological invariant 
employing Berry technology that is
well-suited for semi-metallic bandstructures.
We will demonstrate its
utility through a $2d$ fragile topological model and
a $3d$ Nexus triple-point model.
This also will fill the classification gap for Nexus points 
for a specific set of symmetries. 
However, the invariant is
more general and can be applied in other gap-closing
contexts, including non-electronic ones.
By the
Chern theorem~\cite{Reto_Nguyen_2018}, the Chern number of a closed
manifold has to be an integer. One way to compute the Chern number ($\Chern$) of a 
$2d$ gapped band $|n\rangle$ is by using
Berry connection $\Conn$ and curvature $\Curve$,
\begin{align}
\Chern(n) &=\frac{1}{2\pi}\int_{\vk\in\Surf} \Curve_n\; d^2 k,
\end{align}
with
$\Curve_n =\tfrac{\partial}{\partial k_1}\Conn_2(n)-\tfrac{\partial}{\partial k_2}\Conn_1(n)$ and
$\Conn_i(n) =i\bra{n}\tfrac{\partial}{\partial k_i}\ket{n}.$
If the manifold $\Surf$ is not closed but
has a boundary, then the
Stokes theorem relates the 
curvature integral to the 
connection integral along 
the manifold boundary,
\begin{equation}
\Phase_n=\int_{\vk\in\Surf} \Curve_n d^2 k = \int_{\vk\in\partial\Surf} \Conn(n) \cdot d\vk .
\label{eq:stokes}
\end{equation}
This quantity, in general, does not have to be quantized.
We study the case where our surface of interest is a punctured manifold.
In particular, take a surface in the $3d$ $\BZ$ which crosses one or
more nodal-line band-crossings or a $2d$ $\BZ$ with one or more gap
closing points. 
The relevant bands cannot be uniquely defined where the gap closes.
These surfaces should thus be 
considered as punctured, with a puncture imagined as an infinitesimal boundary circling 
the associated gap-closing point
(Fig.~\ref{fig:puncture_sketch}).
Let us consider first the contribution to Eq.~\ref{eq:stokes}
due to a single puncture involving two bands, say $\tket{n}$ and $\tket{n+1}$.
If this puncture has a linear band-crossing, then the wavefunction $\Psi_{n}(\vk(-s))$
is analytically connected with $\Psi_{n+1}(\vk(s))$ 
for any curve $\{\vk(s)\}\subset\Surf$ 
traversing around the puncture at $s=0$~\cite{Das_Pujari_2020}.
Since $\Psi_{n}(\vk(s))$ is orthogonal to $\Psi_{n+1}(\vk(s))$, 
it follows that
$\Psi_{n}(\vk(s=0^-))$ is orthogonal to $\Psi_{n}(\vk(s=0^+))$, 
i.e. the wavefunction $\Psi_{n}(\vk)$ is orthogonal at antipodal points
along the infinitesimal puncture circling the gap-closing point
(Fig.~\ref{fig:puncture_sketch}).

The $2d$ subspace spanned by the two bands 
has the geometry of a Bloch sphere~\cite{footnote_notation}.
Perpendicular states are represented as antipodal
points on the Bloch sphere.
Thus the image on the Bloch sphere of the
puncture is an antipodal trajectory
(c.f. Fig.~\ref{fig:puncture_sketch}(c)). 
Since any antipodal trajectory necessarily divides the
Bloch sphere into two equal
halves, and the Bloch sphere
includes an integer Berry flux; therefore, the antipodal trajectory divides 
the net Berry flux into two equal halves. This implies that the Berry flux
associated with the antipodal trajectory is half-integer (in units of flux quantum).
That is, \emph{there is a half-integer contribution
to the Chern number on the punctured manifold 
due to a puncture
with a linear band-crossing}. As a corollary, 
a manifold that 
is punctured an odd number of times linearly will 
have a net half-integer Chern number. 
A natural case 
with such structures
are Nexus triple-points to be seen later.

\begin{figure}[t]
\centering
\includegraphics[width=0.95\linewidth]{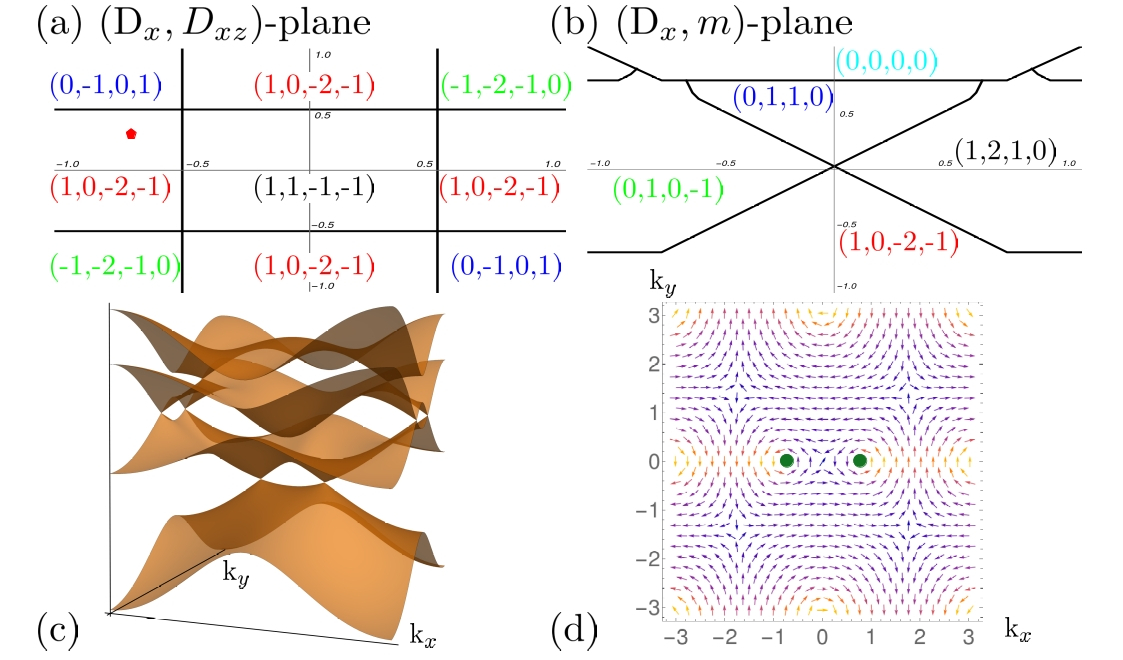}
\caption{
(a) and (b) show the gapless band-topological 
phase diagrams in $\{D_x,D_{xz}\}$ and
$\{D_x,m\}$ planes respectively for Eq.~\ref{eq:Turner_etal}. The other parameters 
are held fixed ($t_1 =1, t_2 =0.5$; $m=-0.5$ for (a);
$D_{xz}=0.5$ for (b)). The transition between the regions (solid lines)
with different punctured-Chern numbers are accompanied by additional gap-closing.
(c) shows the dispersion for a specific
point in the phase diagram marked by a red pentagon 
in (a). (d) shows the
Berry connection map for the bottom band at this point with 
a non-zero punctured-Chern number (the first entry in the corresponding region).}
\label{fig:fragile_topology}
\end{figure}

We first apply this ``punctured-Chern" invariant to a $2d$ fragile
topological model~\cite{Turner_Berg_Stern_2022prl}.
We focus here on capturing the
gapless topology in the parameter space, a big swath of which has 
gap-closing points for all the bands.
The model has four orbitals per unit cell (accounted by
$\sigma$ and $\tau$ Pauli matrices), 
and the Bloch Hamiltonian is
\begin{equation}
H=d_z(\mathbf{k})\sigma_z + d_x(\mathbf{k})\tau_z\sigma_x + d_y(\mathbf{k})\sigma_y + D_x\tau_x + D_{xz}\tau_x \sigma_z
\label{eq:Turner_etal}
\end{equation}
with $d_z(\mathbf{k})=m+t_1 ( \sin^2(k_x/2) + \sin^2(k_y/2) )$, 
$d_x(\mathbf{k})=t_2 \sin(k_x)$, $d_y(\mathbf{k})=t_2 \sin(k_y)$.
The bandstructure at a particular point in the parameter space 
is shown in Fig.~\ref{fig:fragile_topology}(c) with punctured-Chern
invariant $(1,0,-2,-1)$.
Moving around in the parameter space
keeps this invariant unchanged unless one crosses a boundary where 
additional gap-closing occurs
(Fig.~\ref{fig:fragile_topology}(a),(b)).

To visualize the gapless topology, we plot
the Berry connection $\mathbf{\Conn}$ 
for a specific gapless band
in Fig~\ref{fig:fragile_topology}(d). This shows 
the \emph{extra} sources and sinks
of curvature that lead to the non-trivial topology.
In contrast, a trivial punctured-Chern topology
would be the nearest-neighbor hopping model for graphene.
It has no extra sources/sinks of Berry curvature except
the gap-closing points
(punctured-Chern invariant $(0,0)$).
For Eq.~\ref{eq:Turner_etal}, they are 
at high-symmetry points ($(0,\pi)$, $(\pi,0)$, $(\pi,\pi)$ in
Fig~\ref{fig:fragile_topology}(d); also 
$(0,0)$ in other cases) \emph{away} from
the gap-closing points (green dots).
The punctured-Chern invariant captures them
systematically akin to the Chern
invariant for gapped bands. Thus, similar to
quantized anomalous Hall transport~\cite{tknn}, 
punctured-Chern gapless topology 
can drive \emph{non-quantized} anomalous transport (c.f. Eqns.~3,4 of 
Ref.~\onlinecite{Haldane_2004})
constituting a physical effect 
of fragile topological bands~\cite{footnote_stable_speculation}.

\begin{figure}[t]
\centering
\includegraphics[width=0.99\columnwidth]{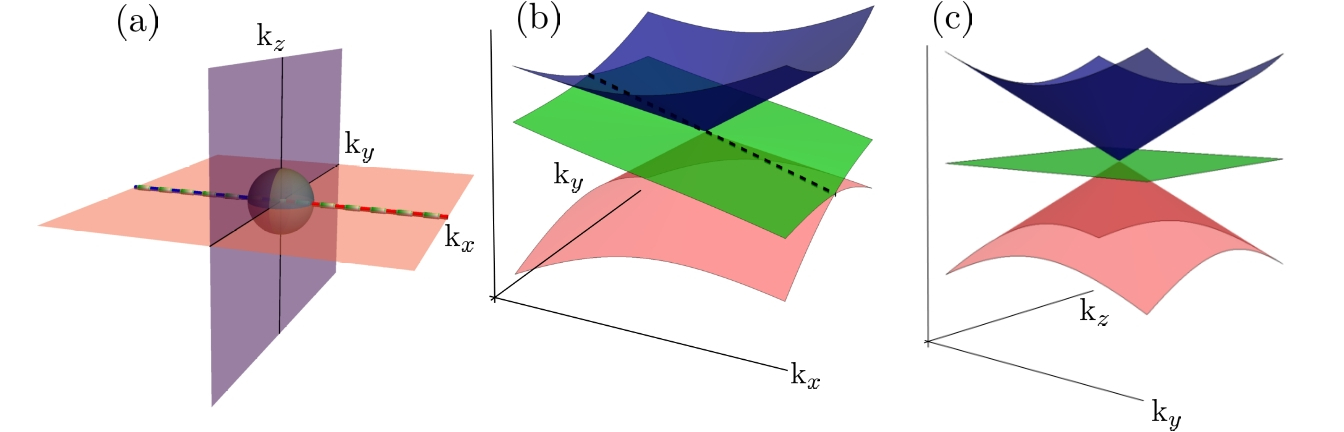}
\caption{
    a $3d$ Nexus triple-point with two
    doubly-degenerate nodal-lines emanating from the triple-point.
    Any surface (e.g. the shown sphere) surrounding the triple-point gets punctured by the nodal lines.
    (b) shows a chosen $2d$ projection containing
    the nodal-lines
    (light red plane in (a)).
    Other projections without the nodal-lines
    (e.g., purple plane in (a))
    look like (c). A three-fold degenerate point that lacks the Nexus structure of (b)
    in any projection is an example of a ``spin-1" chiral fermion.}
\label{fig:nexus_sketch}
\end{figure}

Going now towards $3d$,
we constructively arrive at a $\mathbb{Z} \times \mathbb{Z}$ classification 
for $3d$ triple-point nodes (Fig.~\ref{fig:nexus_sketch}).
The three bands involved in the Nexus point must
have (some multiple of) half-integer punctured-Chern numbers 
in the presence 
of linear band-crossings~\cite{footnote_quad_band_crossing}.  
The sum of the punctured-Chern
numbers over 
the three bands must be an integer,
and 
zero if
filling the three
bands yield the trivial case, which is a natural expectation
in many cases~\cite{footnote_symm_nodal_lines}. 
This implies there are only two independent half-integers.
As half-integers map
to $\mathbb{Z}$, 
therefore the classification of
triple-points is $\mathbb{Z}\times\mathbb{Z}$ 
and computable via the punctured-Chern
invariant.
This invariant is appropriate for a specific choice of internal symmetries. 
Either time reversal or particle-hole may be present, but not both together. 
More generally, we restrict ourselves to the case when \emph{all} the internal symmetries 
map (the vicinity of) the  Nexus point to a different Nexus point, 
implying that there is
\emph{at least one distinct partner} of the Nexus triple-point. 
A rationale for this restriction will be given after discussing
the main results.
This distinct symmetry-related partner
existence (DSPE) condition
then excludes a Nexus point 
at the $\Gamma$ point, 
but does not restrict to only a
single pair of Nexus points. 

The above $\mathbb{Z} \times \mathbb{Z}$ classification result
can be
derived using abstract algebra, which is
conceptually independent of the existence of a constructive invariant.
The general idea is to use exact sequences to calculate the homotopy
groups of the mappings \cite{munkres1974topology,artin2011algebra} that are relevant to
the Nexus (single-particle) band topology. 
A more familiar (many-body) example would be the classes of mappings relevant
for a filled subset of gapped bands mapped onto a general sphere in the sought-after
dimension~\cite{Stone_2009,bernevig2013topological}. Its group structure is achieved
by analyzing the constraints on it
due to elements on either side of the corresponding long exact sequence
\cite{artin2011algebra}. The ``periodic table" 
classification of topological insulators
was similarly achieved using long exact
sequences (e.g., Hopf fibration \cite{Hopf1926,Stone_2009}). A single-particle example
would be a set of non-degenerate single-particle bands, say three bands, whose manifold
structure would be described by the flag manifold $\flag{3} \simeq
\mathrm{U}(3)/\mathrm{U}(1)^3$. For such a manifold, we can conclude $\pi_2(\flag{3}) =
\mathbb{Z} \times \mathbb{Z}$ using the corresponding long-exact sequence by exploiting
its fibration structure as briefly described in Ref.~\onlinecite{supp}.

Here, we have additional structure on the single-particle
Nexus bands due to the emanating nodal-lines.
The presence of gapless points on the manifold
induces the antipodal boundary condition  described previously 
which does not yield a simple fibration structure. 
This prevents directly writing down a long-exact sequence. 
To handle this,
we treat the classification problem using
homotopy pullback~\cite{Carmo2016}, 
and thereby construct a fibration on the space of mappings 
from the vicinity of the triple-point to the band manifold 
and use its long exact sequence 
\cite{artin2011algebra,Frankland_2013,Hatcher2002algebraic} to unravel the classification.
This leads to the $\mathbb{Z}\times  \mathbb{Z}$ homotopy group
for the band manifolds surrounding the Nexus triple-point (for details \cite{supp}).

Next, we discuss a model realization of the above ideas.
To obtain a 
model consistent with DSPE,
we take a $2d$ model of spinless electrons with
three bands and time-reversal symmetry
that has the Nexus structure
with fine-tuned parameters~\cite{Das_Pujari_2019,footnote_su3prb}, 
and build up into $3d$.
To stabilize the triple-point, we consider 
a crystalline system with $C_{3v}$ ditrigonal pyramidal
symmetry and a sublattice anti-symmetry $\Chiral$.
We note that the Type-B class of
Table I in Ref.~\cite{Zhu_etal_2016},
especially those based on 
trigonal space groups (156-161),
can provide promising candidates 
if the Fermi energy lies near the triple-points.
We further assume that chiral symmetry is broken 
(following DSPE), and also broken reflection
symmetry, but the product $\Chiral\sigma_v$ is unbroken.
Thus, the (anti-)symmetries act on the momenta as follows,
\begin{equation}
O_{C_3}=\begin{pmatrix}
\cos\frac{2\pi}{3} & -\sin\frac{2\pi}{3} & 0 \\
\sin\frac{2\pi}{3} & \cos\frac{2\pi}{3} & 0 \\
0 & 0 & 1
\end{pmatrix},\quad 
O_{\Chiral\sigma_v}=\begin{pmatrix}
-1 & 0 & 0 \\
0 & 1 & 0 \\
0 & 0 & 1
\end{pmatrix},
\end{equation}
such that
$gH(\vk)g^{-1}=H(O_g\vk)$,
with $g$ and $O_g$ being 
representations of the symmetry transformation
in the Hilbert space and 
the momentum space respectively.
The following are their actions
on the Hilbert space,
\begin{equation}
C_3 =\begin{pmatrix}
e^{\frac{2\pi i}{3}} & 0 & 0 \\
0 & e^{-\frac{2\pi i}{3}} & 0 \\
0 & 0 & 1
\end{pmatrix}, \qquad
\Chiral\sigma_v =\begin{pmatrix}
0 & 1 & 0 \\
1 & 0 & 0 \\
0 & 0 & 1
\end{pmatrix}.
\label{eq:symHilb}
\end{equation}

For a generic point $\vK$ on the 
three-fold rotational ($C_3$) symmetry axis 
(chosen 
along the
$z$-direction ($\vK = k_z\,\hat{z}$)), 
the most generic Hamiltonian can
be linearized in its vicinity 
($\vp \equiv \vk-\vK$): 
\begin{equation}
H(\vp)=\begin{pmatrix}
\alpha p_z-\epsilon & \beta(p_x+i p_y) & \zeta(p_x-i p_y) \\
\beta(p_x-i p_y) & \epsilon -\alpha p_z & \zeta (p_x+i p_y) \\
\zeta^\ast(p_x+i p_y) & \zeta^\ast(k_p-i p_y) & 0
\end{pmatrix}.
\label{eq:minhil}
\end{equation}
Here, $\alpha$, $\beta$, and $\epsilon$ are 
real parameters,
and $\zeta$ is a 
complex parameter dependent on $\vK$,
i.e. $\alpha \equiv \alpha(\vK)$, $\beta \equiv \beta(\vK)$,
$\zeta \equiv \zeta(\vK)$ and 
$\epsilon \equiv \epsilon(\vK)$.
In the presence of a constant electric field parallel to 
the $C_3$-axis 
given by the vector potential, $\mathbf{A}(t)=Et\,\hat{z}$, 
we get the following time-dependent Hamiltonian after minimal substitution,
\begin{equation}
H(\vp,t)=\begin{pmatrix}
\alpha p_z+\alpha eEt-\epsilon & \beta(p_x+i p_y) & \zeta(p_x-i p_y) \\
\beta(p_x-i p_y) & \epsilon -\alpha p_z - \alpha eEt & \zeta( p_x+i p_y) \\
\zeta^\ast(p_x+i p_y) & \zeta^\ast(p_x-i p_y) & 0
\label{eq:Efield}
\end{pmatrix}.
\end{equation}
Assuming a small 
field,
we 
apply the adiabatic approximation, such
that at time $t=\frac{\epsilon(\vK)}{\alpha e E}$, 
the corresponding point $\vK$ 
would become an
isolated three-fold degenerate point 
(as in Fig.~\ref{fig:nexus_sketch}(c)). 
Since this is true for
any point on the 
$C_3$-axis, some point 
$\vKstar$ may generically also have the property that
$|\beta|=|\zeta|$. Thus at $t^*$ corresponding to
$\vKstar$, a 
triple-point is obtained at $\vKstar$ with 
three emanating \emph{linear} nodal-lines.
It is thus characterized by a half-integer punctured-Chern invariant 
as summarized in Table~\ref{tab:Model}.

\begin{table}[]
\centering
\begin{tabular}{|  c  |  c  |  c  |}
    \hline
    Parameters & Range & puncture-Chern number\\
    \hline\hline
    \multirow{3}{*}{$\beta,\zeta$}
    & \;\;\;\; $\left|\beta\right|<\left|\zeta\right|$ \;\;\;\; &  $\left(-2,0,2\right)$ \\
    \cline{2-3}
    & \;\;\;\; $\left|\beta\right|=\left|\zeta\right|$ \;\;\;\; &  $\left(-1/2,0,1/2\right)$ \\
    \cline{2-3}
    & \;\;\;\; $\left|\beta\right|>\left|\zeta\right|$ \;\;\;\; &  $\left(1,0,-1\right)$ \\
    \hline
\end{tabular}
\caption{
The Chern numbers for the three bands
in three different parameter regimes
for the Hamiltonian of Eq.~\ref{eq:Efield} when the 
$\epsilon$
gets canceled.
For $\left|\beta\right|=\left|\zeta\right|$
with a Nexus triple-point,
we get a half-integer punctured-Chern set
defined on the punctured manifold.
}
\label{tab:Model}
\end{table}

Alternately, 
since the eigenvalues of rotation symmetry representations 
are complex (Eq.~\ref{eq:symHilb} involving $C_3$), this implies 
that the wavefunctions transform as (orbital) angular momentum
$L=1$ 
(also reflected in
the diagonal entries of Eq.~\ref{eq:minhil}). 
We may thus apply a constant magnetic field $B$ 
along the $C_3$-axis,
which will Zeeman couple as $B_z L_z$.
If 
this term were to dominate over minimal coupling terms,
then the Hamiltonian 
becomes essentially as in Eq.~\ref{eq:Efield} with
$\alpha e E t \longrightarrow \mu_0 B$.
Thus, we will again get triple-points 
with half-integer
punctured-Chern numbers when $B=\epsilon(\vKstar)/\mu_0$ (Table~\ref{tab:Model}).
One can understand the half-integer nature 
of the punctured-Chern set as a 
topological transition~\cite{Bouhon_Slager_2022, Unal_Bouhon_Slager_2020} 
between two 
three-fold chiral fermion phases with integer Chern numbers 
that are an odd integer apart~\cite{footnote_stable_topo}.
We can finally write a tight-binding model 
based on the above with the following replacements
\begin{subequations}
\begin{align}
(p_x + i p_y)\longrightarrow &\sum_\ell e^{\frac{2\pi i \ell}{3}}\sin(\va_1\cdot O_{C_3}^{-\ell}\vk)\\
\epsilon  \longrightarrow & \;\; \eta \sum_\ell\cos(\va_1\cdot O_{C_3}^{-\ell}\vk) 
\end{align}
\end{subequations}
involving $C_3$, and
$\alpha p_z \longrightarrow \epsilon(k_z), \; \beta \longrightarrow \beta(k_z),\; \zeta \longrightarrow \zeta(k_z)$.
Here $\ell \in \{1,2,3\}$, $\va_1$ is the primitive lattice vector in the $x$-direction,
$\eta$ is a real parameter, $\epsilon(k_z)$ is a real 
function of $k_z$, and $\beta(k_z)$ and
$\zeta(k_z)$ are respectively real and complex functions 
of $k_z$ whose absolute values
can intersect at some generic momentum
(see Ref. \onlinecite{supp} for 
the explicit formula).

The $\mathbb{Z} \times \mathbb{Z}$ topological invariant delineated 
in this paper can be applied to non-electronic Nexus points in
optical~\cite{Unal_Bouhon_Slager_2020,Park_etal_2021} or phononic~\cite{Li_Song_Jiang_2022}
band-structures. We note here that in other electronic Nexus triple-point 
models~\cite{Zhu_etal_2016,Chang_etal_2017}, 
there are quadratic (instead of linear) band-touchings.
For a quadratic band-touching, the antipodal condition gets lifted
which leads back to a flag manifold of dimension 3.
With DSPE, the classification will thus be $\pi_2(\flag{3})$
which is again $\mathbb{Z} \times \mathbb{Z}$ given by integer Chern sets.
For Ref.~\onlinecite{Chang_etal_2017} and 
\textcolor{blue}{the type-A class of} Ref.~\onlinecite{Zhu_etal_2016}, 
there is a symmetry (time reversal times $c$-axis reflection) which
does not satisfy DSPE.
The Berry connection will thus get further ``constrained" 
in the vicinity of Nexus point. 
This makes the punctured-Chern invariant
inapplicable 
for these cases~\cite{footnote_otherclass}.
It is known that such DSPE-violating symmetries affect the
classification~\cite{Cornfeld_Carmeli_2021} and thereby 
the topological invariants.

We conclude by discussing the relationship between 
the punctured-Chern invariant for Nexus band
topology and physical quantities. 
Because of DSPE, we expect at least 
another ``partner" Nexus point in the $\BZ$. 
Thus we expect that there will be Fermi arcs on the
surface~\cite{footnote_fermi_arc_app}, details depend on
their punctured-Chern numbers.
Due to the half-integral nature from an 
odd number of 
nodal-lines,
appropriate surface cuts will show ``half-integral conductance" 
per cut (in units of $e^2/h$). An equivalent effect
is a half-integral anomalous Hall 
contribution when the chemical potential is nearby a Nexus point~\cite{Haldane_2004}. 
When it
is slightly below (above) the Nexus point,
the Fermi surface 
consists of two hole (electron) Fermi pockets 
that linearly touch each other on the 
nodal-lines. This gives a Hall conductance that 
depends on the 
curvature integral over the two pockets and the 
connection integral to their boundaries
(the infinitesimal loop at the puncture as
in Fig.~\ref{fig:puncture_sketch})
~\cite{footnote_Haldane_2004}. The boundary terms 
cancel out, while the total 
curvature integral will equal that of the bottom (top) sphere, which
is a half-integer~\cite{footnote_integral_Hall}.

The gapless methods involving punctured manifolds 
provide a new perspective and 
calculational tools for topological semi-metals.
In the context of $3d$ Nexus
triple-points, our methods 
go through in presence of certain symmetries as described before. 
This opens up questions regarding
topological invariants to characterize Nexus points 
beyond the DSPE condition. This will
indeed be necessary for several existing 
triple-point semi-metal candidates~\cite{Winkler_Singh_Soluyanov_2019}
including the type-A class in Table I of Ref.~\cite{Zhu_etal_2016}.
More generally, the utility of the punctured-Chern invariant 
lies in the fact that it can naturally capture
semi-metallic
band topology as
seen
for the $2d$ fragile topological model of Eq.~\ref{eq:Turner_etal}.
Another potential application could be to nodal-line semi-metals,
where the punctured-Chern invariant can capture either fragile
or stable topology in the background of the nodal lines in a
$3d$ $\BZ$ similar in spirit to Wilson loop eigenvalue characterization of topological 
insulators~\cite{Alexandradinata_Dai_Bernevig_2014}.

\begin{acknowledgements}
The authors acknowledge fruitful discussions with S.~Carmeli, T.~Holder,
E.~Berg, A.~Stern, J.~S.~Hofmann. AD was supported by the German-Israeli
Foundation Grant No. I-1505-303.10/2019, and CRC 183 (project C01). A.D. also thanks
the Israel planning and budgeting committee (PBC) and the Weizmann Institute of Science,
the Dean of Faculty fellowship, and the Koshland Foundation for financial
support. E.C. was supported by the Deutsche Forschungsgemeinschaft
(DFG, German Research Foundation) project grant 277101999 within the CRC
network TRR 183. 
S.P. acknowledges financial support from SERB-DST, India via grant no. SRG/2019/001419, and in the final stages of writing by grant no. CRG/2021/003024.
\end{acknowledgements}

\bibliographystyle{apsrev}
\bibliography{references}

\begin{thebibliography}{74}
\bibitem[{foo({\natexlab{a}})}]{footnote_notation}
\bibinfo{note}{It is also known as the Riemann sphere, or the complex
  projective line $\CP^1$ in some contexts}.
\bibitem[{foo({\natexlab{b}})}]{footnote_stable_speculation}
\bibinfo{note}{A more speculative question is if there can exist $2d$ models
  that have stable as opposed to fragile topology in presence of gap-closing}.
\bibitem[{foo({\natexlab{c}})}]{footnote_quad_band_crossing}
\bibinfo{note}{In presence of quadratic band-touchings, the bands will have
  integer Chern numbers. One way to think of a quadratic band-touching is as
  two linear band-crossings of the same chirality or winding on top of each
  other}.
\bibitem[{foo({\natexlab{f}})}]{footnote_stable_topo}
\bibinfo{note}{The $2d$ semi-metallic bandstructure surrounding a Nexus triple
  point is also an example of stable topology in presence of gap closing
  ensconced in a $3d$ $\BZ$}.
\bibitem[{foo({\natexlab{j}})}]{footnote_integral_Hall}
\bibinfo{note}{There will also similarly be an integral anomalous Hall
  conductance contribution for the other entries in Table~\ref{tab:Model}}.
\end{thebibliography}

\clearpage

\newpage\newpage



\section*{Supplementary material}

# Supplementry to A $\mathbb{Z} \times \mathbb{Z}$ topological invariant for triple-point Nexus fermions

Ankur Das,[1] Eyal Cornfeld,[1] and Sumiran Pujari[2]

[1]*Department of Condensed Matter Physics, Weizmann Institute of Science, Rehovot 7610001, Israel*
[2]*Department of Physics, Indian Institute of Technology Bombay, Mumbai, MH 400076, India*

Some details connected to the main text are presented here. We discuss the homotopy groups of the different manifolds and specifically the case of the punctured-manifolds associated with the three connected surrounding surfaces for Nexus triple-points. We also present a lattice version of the continuum model with Nexus points discussed in the main text based on the trigonal system, and we show Fermi arcs for a simpler model based on the cubic system.

Below, we present the abstract algebraic proof that the classification of Nexus triple-points with linear band-crossings in presence of the DSPE condition (see main text) is given by $\mathbb{Z} \times \mathbb{Z}$. We start with the simpler case of flag manifolds in Sec. I before going to the case of punctured Nexus manifolds in Sec. II. In Sec. III, we describe a lattice realization of the continuum model with Nexus points discussed in the main text (Eq. 7 of the same). In Sec IV, we show the Fermi arc behavior of a simpler lattice model based on a previous work[1].

## I. HOMOTOPIES OF FLAG MANIFOLDS

We start with the the complex complete flag manifold, denoted by

$$\mathrm{Fl}_3 \stackrel{\mathrm{def}}{=} \mathrm{U}(3)/\mathrm{U}(1)^3 \simeq \mathrm{SU}(3)/\mathrm{U}(1)^2. \tag{1}$$

A fibration, $F \to E \to B$, satisfies[2]

$$\cdots \to \pi_{n+1}(B) \to \pi_n(F) \to \pi_n(E) \to \pi_n(B) \to \pi_{n-1}(F) \to \cdots \tag{2}$$

By $\mathrm{U}(1)^2 \to \mathrm{SU}(3) \to \mathrm{Fl}_3$ we have

$$\begin{array}{ccccccccccc}
\pi_2(\mathrm{SU}(3)) & \to & \pi_2(\mathrm{Fl}_3) & \to & \pi_1(\mathrm{U}(1)^2) & \to & \pi_1(\mathrm{SU}(3)) & \to & \pi_1(\mathrm{Fl}_3) & \to & \pi_0(\mathrm{U}(1)^2) \\
\| & & \| & & \| & & \| & & \| & & \| \\
0 & \to & \mathbb{Z}^2 & \to & \mathbb{Z}^2 & \to & 0 & \to & 0 & \to & 0
\end{array} \tag{3}$$

Here, the dashed arrows are deducted by the structure of the long exact sequence. We conclude that $\pi_1(\mathrm{Fl}_3) = 0$ and $\pi_2(\mathrm{Fl}_3) = \mathbb{Z}^2$.

## II. THE THREE CONNECTED SPHERES

If $Z$ is a connected space then the homotopy pullback[3],

$$\begin{array}{ccc}
 & P & \\
\swarrow & & \searrow \\
X & & Y \\
\searrow & & \swarrow \\
 & Z &
\end{array} \tag{4}$$

is a fibration, $\Omega Z \to P \to X \times Y$, and $P = X \times_Z Y = \{(x, \gamma(t), y) : \gamma(0) = z(x), \gamma(1) = z(y)\}$.

### A. Warmup: two connected spheres

Notice that $\mathrm{Fl}_2 \simeq \mathbb{CP}^1 \simeq S^2$.

The two connected spheres can be viewed as a disk $\partial(D^2) = S^1$. There is a $\mathbb{Z}_2$-action on the boundary sphere which takes each point to its antipodal point. Under this action, the two basis vectors in $\mathrm{Fl}_2$ change places which is the antipodal action on $\mathbb{CP}^1$.

Therefore there is a homotopy pullback

$$\mathrm{Map}^{\mathrm{Conditions}}(D^2, \mathbb{CP}^1) \tag{5}$$

$$\mathrm{Map}(D^2, \mathbb{CP}^1)_{*=\mathrm{hemisphere}} \qquad \mathrm{Map}_{\mathbb{Z}_2}(S^1, \mathbb{CP}^1)$$

$$\mathrm{Map}(S^1, \mathbb{CP}^1)_{*=\mathrm{equator}}$$

such that the number of topologically distinct maps from the disc to $\mathbb{CP}^1$ that obey the antipodality conditions on the boundary are given by $\pi_0(P)$ with $P = \mathrm{Map}^{\mathrm{Conditions}}(D^2, \mathbb{CP}^1)$. The homotopy pullback must be defined relative to a selection of basepoints. For the spaces of maps from the disc to $\mathbb{CP}^1$ we pick as a basepoint the hemisphere map which maps a disc to one of the hemispheres.

The exact sequence reads

$$\begin{array}{ccccccc}
\pi_1(X) \times \pi_1(Y) & \longrightarrow & \pi_0(\Omega Z) & \longrightarrow & \pi_0(P) & \longrightarrow & \pi_0(X) \times \pi_0(Y) \\
\parallel & & \parallel & & \vdots & & \parallel \\
0 \times \mathbb{Z} & \xrightarrow{0} & \mathbb{Z} & \cdots & \{\mathbb{Z}\} & \cdots & 0 \times 0
\end{array} \tag{6}$$

We thus conclude that $\pi_0(P) = \mathbb{Z}$.

*Check:* We know that $\pi_0(P) = \pi_0(\mathrm{Map}^{\mathrm{Conditions}}(D^2, \mathbb{CP}^1)) = \pi_0(\mathrm{Map}^{\mathrm{antipodal}}_{\mathbb{Z}_2}(S^2, \mathbb{CP}^1)) = \{\mathbb{Z}\} - \{2\mathbb{Z}\}$ as labeled by the pushforward $\pi_2(S^2) \to \pi_2(\mathbb{CP}^1)$.

### B. Calculation

The three connected spheres can be viewed as a cylinder $\partial(S^1 \times I) = S^1_u \cup S^1_d$. On each sphere, one of the vectors $v_1 \perp v_2 \perp v_3 \in \mathrm{Fl}_3$ is constant. These vectors are thus points in $\mathbb{CP}^2 \subset \mathrm{Fl}_3$ and they restrict the other pairs to lie in the orthogonal $\mathrm{Fl}_2 \simeq \mathbb{CP}^1 \simeq S^2$. There is a $\mathbb{Z}_2$-action on the boundary spheres which takes each point to its antipodal point. Under this action, the two free basis vectors in $\mathrm{Fl}_3$ change places. That is, for $x_{u,d} \in S^1_{u,d}$ we have

$$x_u; (v_1, v_2, v_3) \mapsto -x_u; (v_2, v_1, v_3), \tag{7}$$
$$x_d; (v_1, v_2, v_3) \mapsto -x_d; (v_1, v_3, v_2). \tag{8}$$

Therefore there is a homotopy pullback

$$\mathrm{Map}^{\mathrm{Conditions}}(S^1 \times I, \mathrm{Fl}_3) \tag{9}$$

$$\mathrm{Map}(S^1 \times I, \mathrm{Fl}_3) \qquad (\mathbb{CP}^2_{v_3} \times \mathrm{Map}_{\mathbb{Z}_2}(S^1_u, \mathbb{CP}^1_{v_1, v_2})) \times (\mathbb{CP}^2_{v_1} \times \mathrm{Map}_{\mathbb{Z}_2}(S^1_d, \mathbb{CP}^1_{v_2, v_3}))$$

$$\mathrm{Map}(S^1_u, \mathrm{Fl}_3) \times \mathrm{Map}(S^1_d, \mathrm{Fl}_3)$$

such that the number of topologically distinct maps from the cylinder to $\mathbb{CP}^2$ that obey the antipodality conditions on the boundary are given by $\pi_0(P)$ with $P = \mathrm{Map}^{\mathrm{Conditions}}(S^1 \times I, \mathrm{Fl}_3)$. Here basepoints are picked akin to the previous section.

The exact sequence reads

$$\begin{array}{ccccccc}
\pi_1(X) \times \pi_1(Y) & \longrightarrow & \pi_0(\Omega Z) & \longrightarrow & \pi_0(P) & \longrightarrow & \pi_0(X) \times \pi_0(Y) \\
\| & & \| & & \vdots & & \| \\
\mathbb{Z}^2 \times ((0 \times \mathbb{Z}) \times (0 \times \mathbb{Z})) & \xrightarrow{\begin{pmatrix} 1&0&0&0 \\ 0&1&0&0 \\ 1&0&0&0 \\ 0&1&0&0 \end{pmatrix}} & \mathbb{Z}^2 \times \mathbb{Z}^2 & \cdots\cdots & \{\mathbb{Z}^2\} & \cdots\cdots & 0 \times 0
\end{array} \qquad (10)$$

Thus, we finally conclude that $\pi_0(P) = \mathbb{Z}^2$.

## III. LATTICE DETAILS

Here we write down the explicit formula for the tight-binding Hamiltonian starting from the continuum model that was described in the main text. It is

$$H(\mathbf{k}) = \begin{pmatrix} \epsilon(k_z) - \eta \sum_\ell \cos(\mathbf{R}_1 \cdot O_{C_3}^{-\ell}\mathbf{k}) & \beta(k_z) \sum_\ell e^{\frac{2\pi i \ell}{3}} \sin(\mathbf{R}_1 \cdot O_{C_3}^{-\ell}\mathbf{k}) & \zeta(k_z) \sum_\ell e^{\frac{2\pi i \ell}{3}} \sin(\mathbf{R}_1 \cdot O_{C_3}^{\ell}\mathbf{k}) \\ \beta(k_z) \sum_\ell e^{\frac{2\pi i \ell}{3}} \sin(\mathbf{R}_1 \cdot O_{C_3}^{\ell}\mathbf{k}) & -\epsilon(k_z) + \eta \sum_\ell \cos(\mathbf{R}_1 \cdot O_{C_3}^{-\ell}\mathbf{k}) & \zeta(k_z) \sum_\ell e^{\frac{2\pi i \ell}{3}} \sin(\mathbf{R}_1 \cdot O_{C_3}^{-\ell}\mathbf{k}) \\ \zeta^*(k_z) \sum_\ell e^{\frac{2\pi i \ell}{3}} \sin(\mathbf{R}_1 \cdot O_{C_3}^{-\ell}\mathbf{k}) & \zeta^*(k_z) \sum_\ell e^{\frac{2\pi i \ell}{3}} \sin(\mathbf{R}_1 \cdot O_{C_3}^{\ell}\mathbf{k}) & 0 \end{pmatrix}, \qquad (11)$$

where $\ell \in \{1, 2, 3\}$, $\mathbf{R}_1$ is the primitive lattice vector in the $r_1$ direction, $\eta$ is a real parameter, $\epsilon(k_z)$ is a real function of $k_z$, and $\beta(k_z)$ and $\zeta(k_z)$ are respectively real and complex functions of $k_z$ whose absolute values intersect at some generic momentum (one may choose one but not both to be constant functions). If one denotes by $k'_z$ the momentum which satisfies $|\beta(k'_z)| < |\zeta(k'_z)|$ the triple-point transition would occur at magnetic fields $B = [3\eta - \epsilon(k'_z)]/\mu$. This model is inspired by those in Ref. 1, specifically the $\Lambda_3$ modification in the $k_z$ direction, however with the $C_3$ symmetry axis lying along $k_z$ as well. In Fig. 1, we show the bulk bandstructure for this model with $|\beta| = |\zeta|$ along the high symmetry lines in the Brillouin zone. Note that 1) the triple point is on the $\Gamma$-$A$ axis as expected as per the discussion in the main text, since it is also the $C_3$-axis as well, and 2) the double degeneracy seen on the $\Gamma$-$K$ axis is actually a consequence of one of the three $C_3$-symmetry related nodal lines which emerge from the triple point intersecting with this axis.

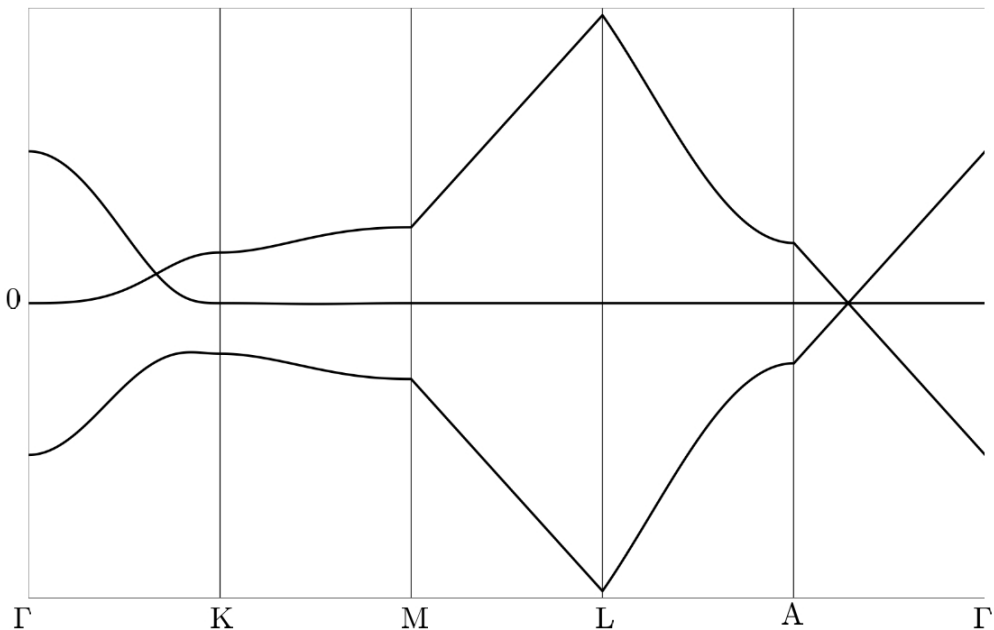

FIG. 1. Bulk energy spectrum for the lattice model in Eq. 11 along the high symmetry lines.

## IV. FERMI ARCS

To illustrate the Fermi arcs due to the topological character of the triple points, we choose a simpler model based on our previous work (Ref. 1):

$$H(\mathbf{p}) = \begin{pmatrix} p_z & p_x - ip_y & p_x - ip_y \\ p_x + ip_y & -p_z & p_x + ip_y \\ p_x + ip_y & p_x - ip_y & 0 \end{pmatrix}. \quad (12)$$

and lift it to a cubic lattice system by the following replacements: $p_z \to \sin k_z$, $p_y \to \sin k_y$, and $p_x \to 2 - \cos k_x - \cos k_y - \cos k_z$. This leads to the Fermi arcs on $(0,0,1)$ plane as seen in Fig. 2.

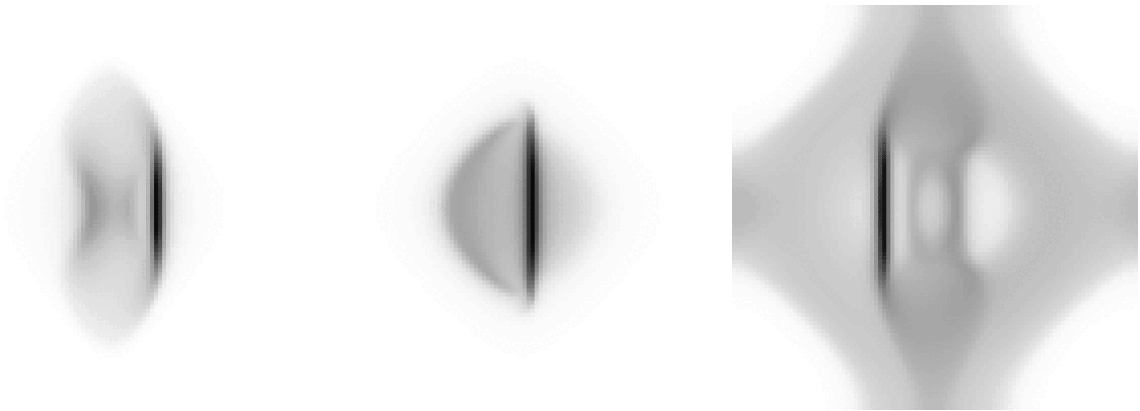

FIG. 2. From left to right, we present equal energy surface state density at three different increasing energies, which shows the Fermi arc structure connecting two pockets.

---

 

\end{document}